\documentclass[sigconf,nonacm]{acmart}

\usepackage{listings}
\usepackage{multirow}
\usepackage{graphics}
\usepackage{hhline}
\usepackage{caption}
\usepackage{subcaption}
\usepackage{listings}
\usepackage{makecell}
\usepackage{fancyhdr}
\usepackage{array}
\usepackage{booktabs}

\begin{document}

\title{Introducing a New Alert Data Set for Multi-Step Attack Analysis}

\author{Max Landauer, Florian Skopik, Markus Wurzenberger}
\email{firstname.lastname@ait.ac.at}
\affiliation{%
  \institution{Austrian Institute of Technology}
  \city{Vienna}
  \country{Austria}
}

\renewcommand{\shortauthors}{Landauer et al.} 

\begin{abstract}
Intrusion detection systems (IDS) reinforce cyber defense by autonomously monitoring various data sources for traces of attacks. However, IDSs are also infamous for frequently raising false positives and alerts that are difficult to interpret without context. This results in high workloads on security operators who need to manually verify all reported alerts, often leading to fatigue and incorrect decisions. To generate more meaningful alerts and alleviate these issues, the research domain focused on multi-step attack analysis proposes approaches for filtering, clustering, and correlating IDS alerts, as well as generation of attack graphs. Unfortunately, existing data sets are outdated, unreliable, narrowly focused, or only suitable for IDS evaluation. Since hardly any suitable benchmark data sets are publicly available, researchers often resort to private data sets that prevent reproducibility of evaluations. We therefore generate a new alert data set that we publish alongside this paper. The data set contains alerts from three distinct IDSs monitoring eight executions of a multi-step attack as well as simulations of normal user behavior. To illustrate the potential of our data set, we experiment with alert prioritization as well as two open-source tools for meta-alert generation and attack graph extraction. 
\end{abstract}


\maketitle

\section{Introduction} \label{intro}

Today's landscape of cyber threats involves more sophisticated tools and complex exploits than ever before. Advanced Persistent Threats (APT) are specifically known to conduct targeted and stealthy attacks that leverage previously unknown attack vectors and are difficult to detect in a timely manner \cite{mtrends2023}. Adversaries such as APTs often progress in similar patterns consisting of several sequential steps that are known as the cyber kill chain, where later stages of attacks generally involve more severe breaches or threats to affected systems and networks \cite{yadav2015technical}. To counteract these threats, security analysts deploy intrusion detection systems (IDS) such as signature-based IDSs that monitor networks for patterns that are known to correspond to malicious activities, as well as anomaly-based IDSs that aim to recognize suspicious deviations from normal user behavior and system utilization \cite{khraisat2019survey}. There is a high diversity of available methods with respect to the monitored data sources (e.g., network packet captures or application log files), operation modes (e.g., expert rules or machine learning), and triggers (e.g., simple string matching or statistical analysis).

A common property of most IDSs is that they are designed to generate alerts for low-level events that are likely the origin of some kind of malicious or undesired actions. However, given the facts that productive systems often generate events in massive amounts and false positive alerts created by possibly unusual yet benign activity can hardly be prevented, the number of alerts generated by IDSs easily becomes overwhelming for security operators and makes manual review and assessment of every single alert infeasible. In fact, studies show that generated alerts may comprise up to 99\% of false positives, causing fatigue and incorrect decision making by operators \cite{alahmadi202299}. In addition, low priority alerts that occur in high volumes (e.g., resulting from basic scanning) could conceal more relevant alerts that occur simultaneously but only in small numbers.

It is thus necessary to prioritize or weight alerts and enrich them with contextual information such as their relation to each other in the view of a larger attack chain. In academic research, this task is referred to as multi-step attack analysis and aims at the aggregation or correlation of single alerts into higher-level abstractions of attack scenarios. However, recent studies indicate that this objective is difficult to achieve for several reasons. In particular, each step of the attack chain often generates multiple alerts, for example, when malicious activities leave detectable traces in multiple monitored sources; at the same time, the same or similar alerts may be generated as part of separate attack steps or even entirely different attack scenarios \cite{landauer2022dealing}. Thereby, real-world attacks often involve multiple systems within the same network, causing that analysis of isolated machines only provides an incomplete view on the attack chains and necessitate to combine relevant traces across several distributed data sources \cite{levshun2023survey}. Moreover, multi-step attack analysis does not only require to infer the nature of each step, but also the links between them \cite{navarro2018systematic}. Even though multi-step attacks generally follow kill chains, inferring these links is especially difficult since there is not necessarily a direct mapping between them, e.g., there may be arbitrary many steps from the same kill chain stage involved \cite{zhou2021detecting}. Even worse, there are situations where some attack steps appear normal in isolation and can only be identified when also other steps are considered \cite{chadza2020analysis}. 

There is a need to address these problems with novel scientific approaches; however, as pointed out in several recent surveys \cite{navarro2018systematic, kotenko2022systematic, husak2018survey, levshun2023survey}, one of the main issues holding back the research community is the lack of publicly available data sets for experimentation and evaluation. Existing public data sets origin from outdated or oversimplified systems \cite{husak2018survey, kotenko2022systematic}, only consider a single source of data \cite{chadza2020analysis}, and fit the purpose of intrusion detection rather than multi-step attack analysis \cite{husak2018survey, landauer2022dealing}. As a consequence, researchers often resort to private data sets from productive systems that prevent reproducibility and comparability of results \cite{kotenko2022systematic}.

Alongside this paper we therefore publish a new alert data set that aims to resolve this gap. We ensure that the properties of our data set align with several of the challenges inherent to the problem domain of multi-step attack analysis, including high volumes of alerts and false positives \cite{navarro2018systematic}, the presence of alerts from heterogeneous IDSs involving diverse detection techniques and alert formats \cite{navarro2018systematic}, collection of alerts from multiple network components and data sources \cite{levshun2023survey, chadza2020analysis}, inclusion of anomaly-based alerts that lack contextual information or direct connections to root causes \cite{navarro2018systematic}, changes of attack step order and attack parameters \cite{zhou2021detecting, navarro2018systematic}, and a clear and repeatable attack plan \cite{husak2018survey}. To this end we select the synthetically generated and publicly available Austrian Institute of Technology Log Data Set version 2 (AIT-LDSv2) \cite{landauer2023maintainable} for alert generation, because it provides system log data and network packet captures of simulated normal behavior and a multi-step attack executed with variations of attack parameters in eight different environments. In particular, we forensically analyze the AIT-LDSv2 with Wazuh IDS as well as AMiner IDS and collect alerts from Suricata IDS to generate our data set. To the best of our knowledge, this is the first public data set specifically designed to enable researchers to evaluate approaches for multi-step attack analysis, including related research areas focusing on prioritization, filtering, aggregation, and correlation of alerts, as well as the generation of meta-alerts, handling of false positives, and more. We publish the alert data set on an openly accessible data sharing platform\footnote{AIT-ADS available at \url{https://zenodo.org/record/8263181} (accessed 2023-08-24)} and also provide the scripts for reproduction and analysis of the data set\footnote{GitHub repository available at \url{https://github.com/ait-aecid/alert-data-set} (accessed 2023-08-24)}. We summarize our contributions as follows:

\begin{itemize}
	\item A new public data set comprising alerts generated by multiple detectors monitoring diverse data sources for multi-step attack scenarios independently executed in eight networks,
	\item an analysis of the generated data set for the purpose of alert filtering and detector prioritization, and
	\item an illustrative application of two open-source tools for alert aggregation and attack graph mining.
\end{itemize}

The remainder of the paper is structured as follows. Section \ref{related} reviews existing alert data sets and their applications. We then outline our approach of generating the data set in Sect. \ref{dataset} and emphasize relevant characteristics of the data. In Sect. \ref{prioritization}, we describe a method to weight detectors and filter irrelevant alerts. Using the filtered data set, Sect. \ref{analysis} then illustrates the application of two open-source frameworks for alert aggregation and attack graph generation. We discuss our findings in Sect. \ref{discussion}. Finally, Sect. \ref{conclusion} concludes the paper. 

\section{Background \& Related Work} \label{related}

When it comes to scientific evaluations of approaches for multi-step attack analysis, availability of appropriate data sets is a critical factor. Unfortunately, surveys have shown that around 40\%-50\% of works published in this research field rely on private data sets, meaning that their results are not reproducible \cite{navarro2018systematic, kotenko2022systematic, levshun2023survey}. Publishing these data sets is often not permitted due to the fact that they origin from productive environments and possibly involve sensitive data. Moreover, more than half of the remaining publications that do use public data sets resort to data sets from the DARPA collection \cite{husak2018survey}. These data sets have been heavily criticized for a multitude of reasons, such as being outdated (they were generated in the years 1998-2000) and apparent oversimplification of both background traffic and attack manifestations \cite{thomas2008usefulness}. The data sets also only involve network traffic and are thus not suitable to generate alerts from host-based IDSs that analyze system log data \cite{chadza2020analysis}. 

One of the main problems is that publicly available data sets are originally not intended to be used for multi-step attack analysis; they are primarily designed as data sets for IDS research that researchers resort to for evaluation of multi-step attack analysis methods for the lack of better alternatives \cite{husak2018survey}. For example, Chadza et al. run Snort IDS on one of the DARPA 2000 scenarios to obtain an alert data set, which they use to develop and evaluate an approach to detect the current stage of the attack and predict the upcoming ones. Specifically, they leverage Hidden Markov Models and compare various training and initialization algorithms. 

Landauer et al. \cite{landauer2022dealing} applies signature- and anomaly-based IDSs on the AIT-LDSv1 \cite{landauer2020have}, which comprises log data from four multi-step attacks. While this is an adequate setting for evaluation of their aggregation algorithm to generate meta-alerts across systems, the AIT-LDSv1 only comprises alerts from a single server rather than multiple components in the network. For this reason, we select the AIT-LDSv2 that improves upon this issue among several others, including extensiveness of simulations for normal behavior, realism of network layout, collection of network traffic, and reliability of attack labels. In Sect. \ref{aggregation} we test their approach for alert aggregation on our newly generated data set and describe the results.

Another source of alert data sets is provided by the annual Collegiate Penetration Testing Competition\footnote{Collegiate Penetration Testing Competition, \url{https://cp.tc/} (accessed 2023-08-10)} (CPTC). The organizers of the event provide student teams with simulated environments of corporate or industrial networks and task them to discover vulnerabilities of software services for educational purposes. In course of this event, log data and alerts from IDSs deployed in the networks are collected and published as publicly available data sets, which have since been used by several researchers. For example, Perry et al. \cite{perry2018differentiating} use the CPTC-2017, which comprises alerts from Suricata IDS triggered by the attacks from ten student teams, to train an LSTM and predict upcoming attack stages. While the CPTC data sets appear highly useful for analysis of attack strategies, it is important to take into account that these data sets primarily resemble comprehensive penetration tests rather than targeted attacks since students are tasked to discover as many vulnerabilities as possible \cite{moskal2018extracting, meyers2022examining}. Close examination of the CPTC-2017 has also shown that attack sequences commonly used across multiple teams mostly relate to scans or other basic attacks, while more advanced attack techniques are usually unique to one or few teams \cite{moskal2018extracting}. In-depth analysis of the CPTC-2019 also showed that there is only little overlap of the discovered vulnerabilities across different teams \cite{meyers2022examining}. As a consequence, single alert occurrences need to be mapped to some high-level categories of attack stages in order to extract common attack patterns of multi-step attacks from the data sets. This is accomplished by Nadeem et al. \cite{nadeem2021sage}, who map alert signatures from Suricata IDS to the abstract categories of the Action-Intent Framework \cite{moskal2020framework} for the purpose of generating attack graphs. We discuss their approach in more detail in Sect. \ref{graph} and use it to generate an attack graph for the data set introduced in this paper.

Beside the DARPA data set, Zhou et al. \cite{zhou2021detecting} use the ISCXIDS2012 \cite{shiravi2012toward} and the CIC-IDS2017 \cite{sharafaldin2018toward} for evaluation of their approach to detect multi-step attacks in sequences of alerts. Both data sets only comprise network traffic and are primarily designed for IDS research. While the ISCXIDS2012 already contains traces of multi-step attacks, the CIC-IDS2017 only involves individual attack steps that are rearranged by the authors to fit their purpose. To evaluate their approach, Ben et al. \cite{ben2020cybersecurity} use the CTF data set from DEFCON\footnote{DEFCON, \url{https://defcon.org/} (accessed 2023-08-10)} 2017, which contains alerts from the network-based Snort IDS with a total of 36 unique alert signatures. They experiment with deep neural networks for the prediction of attack types based on involved IP addresses and previously observed attack types.

Ramaki et al. \cite{ramaki2021towards} use the Scan-of-the-Month data set provided by the Honeynet project\footnote{Honeynet project, \url{https://honeynet.onofri.org/scans/index.html} (accessed 2023-08-10)} to evaluate their approaches for alert filtering, similarity-based clustering, and cluster summarization. The advantage of this data set is that it comprises logs from heterogeneous sources, including a network-based IDS, firewall events, and system logs. Another honeypot data set is presented by Sperotto et al. \cite{sperottoIPOM2009}, who collect network flows and label them manually. Hus{\'a}k et al. \cite{husak2020dataset} also provide a honeypot data set from an alert sharing platform monitored by network-based IDSs. Common problems with data sets from honeypots include lack of control over attacker activities and difficulties in labeling unknown traffic \cite{landauer2023maintainable}. 

\section{Alert Data Set} \label{dataset}

This section describes the alert data set published alongside this paper. We first explain how we generated the data set and then highlight relevant characteristics.

\subsection{Generation} \label{generation}

This section describes the original log data set as well as three open-source IDSs used to generate the alert data set.

\subsubsection{AIT-LDSv2} \label{aitlds}

Publicly available alert data sets are scarce, but so are suitable log data sets containing traces of attacks \cite{landauer2023maintainable}. One of them is the AIT-LDSv2\footnote{AIT-LDSv2, \url{https://zenodo.org/record/5789064} (accessed 2023-08-10)} published in 2022. The data set contains synthetic network traffic and system logs collected from a virtual test environment that represents an enterprise network, which consists of an intranet zone with a file share and an intranet server, a demilitarized zone with a VPN server, a mail server, and a cloud storage, and an Internet zone with a DNS server and additional mail servers, all connected through a firewall. Normal behavior is generated by extensive state machines that simulate employees interacting with available services. The simulation was set up to run for multiple days and involves two main attack cases. The first one is a multi-step attack comprising several scans using the tools Nmap for service and host scans, Dirb for directory scans, and WPScan for scanning the intranet server running a WordPress platform, as well as exploits to upload a webshell through a vulnerable WordPress plugin, password cracking, installation of a reverse shell, and privilege escalation. The second attack case exfiltrates sensitive data from the file share over stealthy DNS requests using the tool DNSteal. As a challenge to anomaly-based IDSs, the exfiltration case was designed to be already active at the beginning of the data set and stop at a specific point in time, which is more difficult to detect than a new service starting. We refer to the publication by Landauer et al. \cite{landauer2023maintainable} for a detailed description of the log data set. 

The AIT-LDSv2 has some unique features that, to the best of our knowledge, make it the only publicly available data set suitable to obtain alerts that adequately address common challenges of multi-step attack analysis (cf. Sect. \ref{intro}). First, the data set comprises eight scenarios named \textit{fox}, \textit{harrison}, \textit{russellmitchell}, \textit{santos}, \textit{shaw}, \textit{wardbeck}, \textit{wheeler}, and \textit{wilson}, that all involve the same environment and attack cases. However, each scenario was designed to have unique variations regarding the attack cases (i.e., attack parameters such as scan intensities), environment (e.g., number of deployed servers), and user simulations (e.g., roles of employees and their individual preferences). Alerts generated from these eight instances thus enable derivation of training and test data sets for attack step prediction, evaluation of approaches for alert aggregation across organizations, computation of similarities for single or combined attack steps, etc. Second, the data set comprises packet captures from network traffic as well as log files from various sources, such as low-level Audit logs, Apache access logs, DNS logs, syslog, CPU logs, and several application logs. This means that we are able to generate alerts from several sources using IDSs that operate on network as well as host level. Third, log data is collected from every component in the network. As attacks leave traces on multiple machines, generated alerts need to be correlated accordingly. Fourth, a large portion of the data is collected during normal operation. Generating alerts from that data yields false positives, which enable evaluation of alert prioritization and filtering mechanisms. Fifth, the data set is fully labeled. We are therefore able to assign labels for attack phases also to alerts based on their occurrence time.

\subsubsection{Intrusion Detection Systems} \label{ids}

We obtain alerts from three open-source IDSs by forensically processing the AIT-LDSv2 with Wazuh and AMiner and collecting alerts from Suricata. Note that even though all generated alerts are in JSON format, there is no common schema for the alert objects since different detectors use specific fields. For example, detector signatures reside in the ``description'' field of Wazuh alerts and the ``AnalysisComponentName'' field of AMiner alerts. Table \ref{tab:alerts} summarizes all 93 unique detector signatures (34 from AMiner, 29 from Suricata, and 30 from Wazuh). In the following, we refer to these detectors by the abbreviations shown in the table, where the first token indicates the IDS (AMiner - \textit{A}, Suricata - \textit{S}, Wazuh - \textit{W}), the second token refers to the log source or data field where the alert was found (Apache access - \textit{Acc}, Audit - \textit{Aud}, authentication logs - \textit{Aut}, Apache error - \textit{Err}, DNS - \textit{Dns}, mail logs - \textit{Mai}, Syslog - \textit{Sys}, resource monitoring - \textit{Mon}, packet captures - \textit{Dns/Flw/Htt/Nat/Smt/Tls}, multiple sources - \textit{All}), and the third token is event-specific. For simplicity, we use the same abbreviations for some signatures with similar implications. The following paragraphs briefly describe each deployed IDS. 

\begin{table}
	\renewcommand{\arraystretch}{0.65}
	\tiny
	\caption{Detectors in the data set and abbreviations}
	\label{tab:alerts}
	\begin{tabular}{ll}
		\textbf{Detector} & \textbf{Abbreviation} \\ \hline
		New characters in Apache Access referer. & A-Acc-Chr1 \\ \hline
		New characters in Apache Access request. & A-Acc-Chr2 \\ \hline
		Unusual occurrence frequencies of Apache Access request methods. & A-Acc-Clc \\ \hline
		High entropy in Apache Access referer. & A-Acc-Ent1 \\ \hline
		High entropy in Apache Access request. & A-Acc-Ent2 \\ \hline
		High entropy in Apache Access user agent. & A-Acc-Ent3 \\ \hline
		Unusual occurrence frequencies of Apache Access logs. & A-Acc-Frq \\ \hline
		New request method in Apache Access log. & A-Acc-Val1 \\ \hline
		New status code in Apache Access log. & A-Acc-Val2 \\ \hline
		New event type. & A-All-Evt \\ \hline
		New apparmor parameter combination in Audit logs. & A-Aud-Com1 \\ \hline
		New cred\_acq parameter combination in Audit logs. & \multirow{3}{*}[-1pt]{A-Aud-Com2} \\ \cline{1-1}
		New cred\_disp parameter combination in Audit logs. &  \\ \cline{1-1}
		New cred\_refr parameter combination in Audit logs. &  \\ \hline
		New login parameter combination in Audit logs. & A-Aud-Com3 \\ \hline
		New service\_start parameter combination in Audit logs. & \multirow{2}{*}{A-Aud-Com4} \\ \cline{1-1}
		New service\_stop parameter combination in Audit logs. &  \\ \hline
		New syscall parameter combination in Audit logs. & A-Aud-Com5 \\ \hline
		New user\_acct parameter combination in Audit logs. & \multirow{6}{*}[-2pt]{A-Aud-Com6} \\ \cline{1-1}
		New user\_auth parameter combination in Audit logs. &  \\ \cline{1-1}
		New user\_cmd parameter combination in Audit logs. &  \\ \cline{1-1}
		New user\_end parameter combination in Audit logs. &  \\ \cline{1-1}
		New user\_login parameter combination in Audit logs. &  \\ \cline{1-1}
		New user\_start parameter combination in Audit logs. &  \\ \hline
		Unusual occurrence frequencies of DNS log events. & A-Dns-Clc1 \\ \hline
		Unusual occurrence frequencies of DNS query IPs. & A-Dns-Clc2 \\ \hline
		Unusual occurrence frequencies of DNS query records. & A-Dns-Clc3 \\ \hline
		New characters in DNS domain. & A-Dns-Chr \\ \hline
		High entropy in DNS domain. & A-Dns-Ent \\ \hline
		Unusual occurrence frequencies of query records in DNS logs. & A-Dns-Frq \\ \hline
		New ip address in DNS logs. & A-Dns-Val1 \\ \hline
		New query record in DNS logs. & A-Dns-Val2 \\ \hline
		CPU value deviates from average in monitoring logs. & A-Mon-Avg \\ \hline
		CPU value out of expected range in monitoring logs. & A-Mon-Rng \\ \hline
		ET INFO Suspicious Domain (*.ga) in TLS SNI & S-Dns-Dom \\ \hline
		ET DNS DNS Lookup for localhost.DOMAIN.TLD & S-Dns-Loo \\ \hline
		SURICATA DNS Unsolicited response & S-Dns-Uns \\ \hline
		ET DNS Query for .cc TLD & \multirow{4}{*}[-1pt]{S-Dns-Qry1} \\ \cline{1-1}
		ET DNS Query for .su TLD (Soviet Union) Often Malware Related &  \\ \cline{1-1}
		ET DNS Query for .to TLD &  \\ \cline{1-1}
		ET DNS Query to a *.pw domain - Likely Hostile &  \\ \hline
		ET INFO DNS Query for Suspicious .ga Domain & S-Dns-Qry2 \\ \hline
		ET INFO Observed DNS Query to .biz TLD & S-Dns-Qry3 \\ \hline
		ET INFO Observed DNS Query to .cloud TLD & S-Dns-Qry4 \\ \hline
		ET SCAN Behavioral Unusual Port 445 traffic Potential Scan or Infection & S-Flw-445 \\ \hline
		ET POLICY GNU/Linux APT User-Agent Outbound & \multirow{2}{*}{S-Flw-Apt} \\
		likely related to package management & \\ \hline
		ET HUNTING Possible COVID-19 Domain in SSL Certificate M2 & \multirow{3}{*}[-1pt]{S-Flw-Cov} \\ \cline{1-1}
		ET HUNTING Suspicious Domain Request for Possible COVID-19 Domain M1 &  \\ \cline{1-1}
		ET HUNTING Suspicious TLS SNI Request for Possible COVID-19 Domain M1 &  \\ \hline
		ET SCAN Possible Nmap User-Agent Observed & S-Flw-Nmp \\ \hline
		SURICATA HTTP gzip decompression failed & S-Htt-Gzp \\ \hline
		SURICATA HTTP unable to match response to request & S-Htt-Mat \\ \hline
		SURICATA HTTP invalid response chunk len & S-Htt-Res \\ \hline
		ET INFO Session Traversal Utilities for NAT (STUN Binding Request) & \multirow{2}{*}{S-Nat-Trv} \\ \cline{1-1}
		ET INFO Session Traversal Utilities for NAT (STUN Binding Response) &  \\ \hline
		SURICATA SMTP invalid reply & S-Smt-Rep \\ \hline
		SURICATA SMTP no server welcome message & S-Smt-Wel \\ \hline
		SURICATA TLS certificate invalid der & S-Tls-Crt \\ \hline
		ET INFO TLS Handshake Failure & S-Tls-Fai \\ \hline
		SURICATA TLS invalid handshake message & S-Tls-Hnd \\ \hline
		SURICATA TLS invalid record/traffic & S-Tls-Rec \\ \hline
		SURICATA TLS invalid SSLv2 header & S-Tls-Ssl \\ \hline
		SURICATA TLS invalid record type & S-Tls-Typ \\ \hline
		Web server 400 error code. & W-Acc-400 \\ \hline
		Web server 500 error code (Internal Error). & W-Acc-500 \\ \hline
		Common web attack. & W-Acc-Att \\ \hline
		CMS (WordPress or Joomla) brute force attempt. & W-Acc-Brt \\ \hline
		CMS (WordPress or Joomla) login attempt. & W-Acc-Cms \\ \hline
		Suspicious URL access. & W-Acc-Sus \\ \hline
		IDS event. & W-All-Evt \\ \hline
		First time this IDS alert is generated. & W-All-Ids \\ \hline
		Multiple IDS alerts for same id (ignoring now this id). & \multirow{2}{*}{W-All-Mul1} \\ \cline{1-1}
		Multiple IDS alerts for same id. &  \\ \hline
		Multiple IDS events from same source ip. & \multirow{2}{*}{W-All-Mul2} \\ \cline{1-1}
		Multiple IDS events from same source ip (ignoring now this srcip and id). &  \\ \hline
		Multiple web server 400 error codes from same source ip. & W-All-Mul3 \\ \hline
		Auditd: SELinux permission check. & W-Aud-Sel \\ \hline
		PAM: Login session closed. & \multirow{2}{*}{W-Aut-Pam1} \\ \cline{1-1}
		PAM: Login session opened. & \\ \hline
		PAM: User login failed. & W-Aut-Pam2 \\ \hline
		PAM: Multiple failed logins in a small period of time. & W-Aut-Pam3 \\ \hline
		sshd: authentication success. & W-Aut-Ssh1 \\ \hline
		sshd: insecure connection attempt (scan). & W-Aut-Ssh2 \\ \hline
		First time user executed sudo. & \multirow{2}{*}{W-Aut-Sud} \\ \cline{1-1}
		Successful sudo to ROOT executed. &  \\ \hline
		User successfully changed UID. & W-Aut-Uid \\ \hline
		Apache: Attempt to access forbidden directory index. & W-Err-Fbd1 \\ \hline
		Apache: Attempt to access forbidden file or directory. & W-Err-Fbd2 \\ \hline
		Dovecot brute force attack (multiple auth failures). & W-Mai-Brt \\ \hline
		Dovecot Invalid User Login Attempt. & W-Mai-Inv \\ \hline
		ClamAV database update & W-Sys-Cav \\ \hline
		Dovecot Authentication Success. & W-Sys-Dov \\ \hline
		syslog: User authentication failure. & W-Sys-Fai \\ \hline
	\end{tabular}
\vspace{-7pt}
\end{table}

\textbf{Wazuh}\footnote{Wazuh, \url{https://wazuh.com/} (accessed 2023-08-10)} is a host-based and signature-based IDS that scans log files for potentially malicious events. Wazuh relies on a data base of expert rules that specify textual patterns that need to match in log lines to trigger alerts. There are also advanced rules that are only active when other rules have been triggered beforehand or some patterns have matched a minimum amount of times in a specific time interval. Since Wazuh does not support forensic analysis of log files, we created a script that reads out the timestamps of log files in the AIT-LDSv2 and feeds them into Wazuh in real-time. 

\textbf{Suricata}\footnote{Suricata, \url{https://suricata.io/} (accessed 2023-08-10)} is a network-based and signature-based IDS that may also be used as an intrusion prevention system (IPS). Suricata inspects network packets and conducts pattern matching using a data base of expert rules similar to Wazuh. Specifically, Suricata matches flows by protocol, IP addresses, ports, etc. The authors of the AIT-LDSv2 already deployed Suricata on the servers in the network. Accordingly, Suricata alerts are already available in the data set and can be conveniently collected by Wazuh.

\textbf{AMiner}\footnote{AMiner, \url{https://github.com/ait-aecid/logdata-anomaly-miner} (accessed 2023-08-10)} is a host-based and - contrary to Wazuh and Suricata - anomaly-based IDS. This means that it is necessary to train AMiner with sufficiently many logs corresponding to normal system behavior so that the models used for detecting deviations adequately represent normal activities. We therefore use the first two days of each scenario in the AIT-LDSv2 for training and switch the AMiner to detection mode afterwards, so that the learned models are not affected by the attacks. Moreover, there is no default configuration for AMiner that works out-of-the-box; instead, we empirically select and configure the following detectors: (i) event detection (\textit{Evt}) detects new event types that have not been observed before, (ii) value detection (\textit{Val}) detects new categorical event parameters, (iii) combo detection (\textit{Com}) is similar to value detection but works on combinations of event parameters, (iv) character detection (\textit{Chr}) recognizes new characters in textual parameters, (v) entropy detection (\textit{Ent}) analyzes likelihoods of character transitions in textual parameters, (vi) frequency detection (\textit{Frq}) applies seasonal time-series forecasting on event frequencies, (vii) count detection (\textit{Clc}) detects unusual event count distributions in time windows, (viii) range detection (\textit{Rng}) detects numeric parameters outside of learned minimum and maximum bounds, and (ix) average change detection (\textit{Avg}) analyzes numeric parameters for deviating means and variances. For more details on these detection mechanisms, we refer to the AMiner paper \cite{landauer2023aminer} and our repository (see link in Sect. \ref{intro}).

\subsection{Characteristics}

This section highlights relevant characteristics of the alert data set, specifically patterns and frequencies of alert occurrences.

\subsubsection{Scenario Timelines} \label{timelines}

\begin{figure*}
	\centering
	\includegraphics[width=1\textwidth]{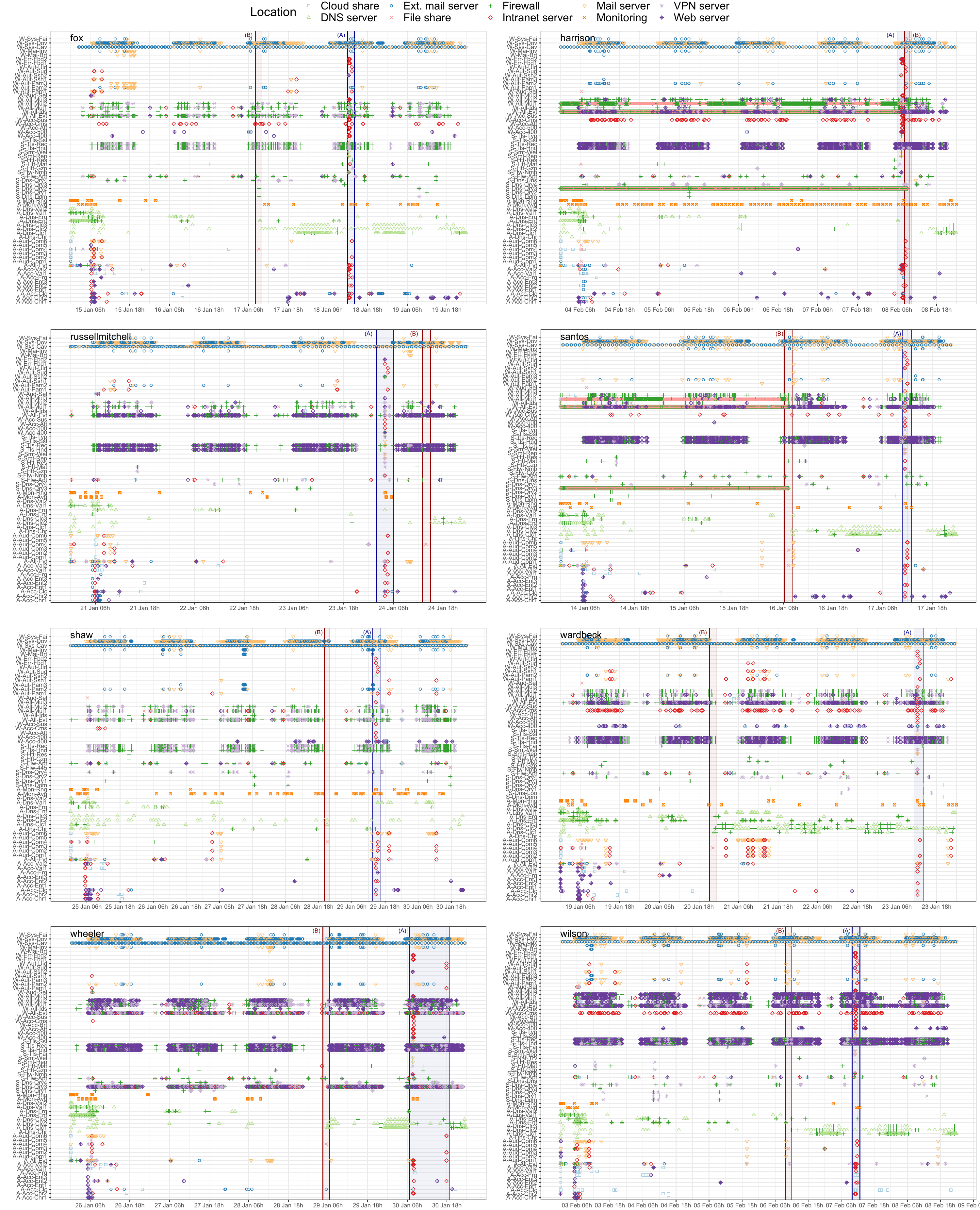}
	\caption{Timelines of alert type occurrences. Shaded intervals indicate multi-step attack (A/blue) and data exfiltration (B/red).}
	\label{fig:datasets}
\end{figure*}

Our generated alert data set comprises alerts from the three aforementioned IDS applied on each of the eight scenarios provided in the AIT-LDSv2. Figure \ref{fig:datasets} plots the alerts generated from each detector during the entire time span of every scenario, including phases of normal activity. Thereby, each alert occurrence is marked by a distinct symbol and color to differentiate the location of detection, i.e., the network component where the IDS reported the alert. The time windows where the two attack cases of the (A) multi-step attack and (B) data exfiltration leave detectable traces in the logs are indicated by shaded intervals of blue and red colors respectively.

Examining these plots makes it immediately clear that several detectors report a high number of false positives, i.e., alerts occurring outside of the attack time windows. We point out that referring to these alerts as false positives may be misleading; the detectors correctly report these events as expected, it is just the case that the events themselves do not correspond to any activities related to the attacks in the context of these scenarios. For example, alerts are triggered when the ClamAV service attempts to update, which occurs roughly once every hour on multiple components (\textit{W-Sys-Cav}). Other detectors are triggered by normal user activity and thus only occur during daytime, for example, users logging into their mail accounts generate alerts that notify on a successful authentication (\textit{W-Sys-Dov}). Another interesting observation is that almost all AMiner detectors report multiple false positives in the first half of the first day of each scenario, which is the result of training the models (e.g., adding new categorical values to the value detector) that are still incomplete and not representative for the system behavior at this point. As visible in the plots, the frequencies of these false positives quickly diminish for most detectors and there are hardly any false positives from the second day onward.

The plots also show that the multi-step attack triggers several alerts from each of the three IDSs in every scenario. As expected, most of the alerts stem from the intranet server, which is the main target of this attack case. We present a more detailed view on the multi-step attack in Sect. \ref{prioritization}. The data exfiltration attack also triggers several alerts. In every scenario, stopping the exfiltration service is detected as anomalous by the AMiner (\textit{A-Aud-Com4}) as this behavior has not been observed in the training phase. Moreover, since the exfiltration generates a high number of events while it is active, the AMiner is able to recognize deviations of DNS event frequencies once the service stops and reports alerts until the end of the simulations (\textit{A-Dns-Clc1/2/3}). In the \textit{harrison} and \textit{santos} scenarios, the exfiltration itself is also detected by Suricata, which produces a high number of alerts until the service is stopped (\textit{S-Dns-Qry3}). The reason for this is that only in these scenarios the domain of the attacker has a ``.biz'' top-level-domain, which is considered suspicious. These Suricata alerts also trigger Wazuh rules, creating additional alerts (\textit{W-All-Evt} and \textit{W-All-Mul1}). 

\subsubsection{Alert counts in scenarios} \label{counts}

Across all scenarios, the total number of alerts is 2,655,821, where 2,293,628 (86.4\%) origin from Wazuh, 306,635 (11.5\%) from Suricata, and 55,558 (2.1\%) from AMiner. However, alert counts vary strongly depending on the scenario, e.g., Wazuh generates 560,265 alerts in \textit{wilson} but only 32,302 in \textit{russellmitchell}. The reasons for that are manifold, but mostly depend on the length of the simulation (6 days vs 4 days), the number of employees simulated as part of the scenario (24 vs 10), and the parameters of attack executions (extensive vs basic scanning).

\begin{figure*}
	\centering
	\includegraphics[width=1\textwidth]{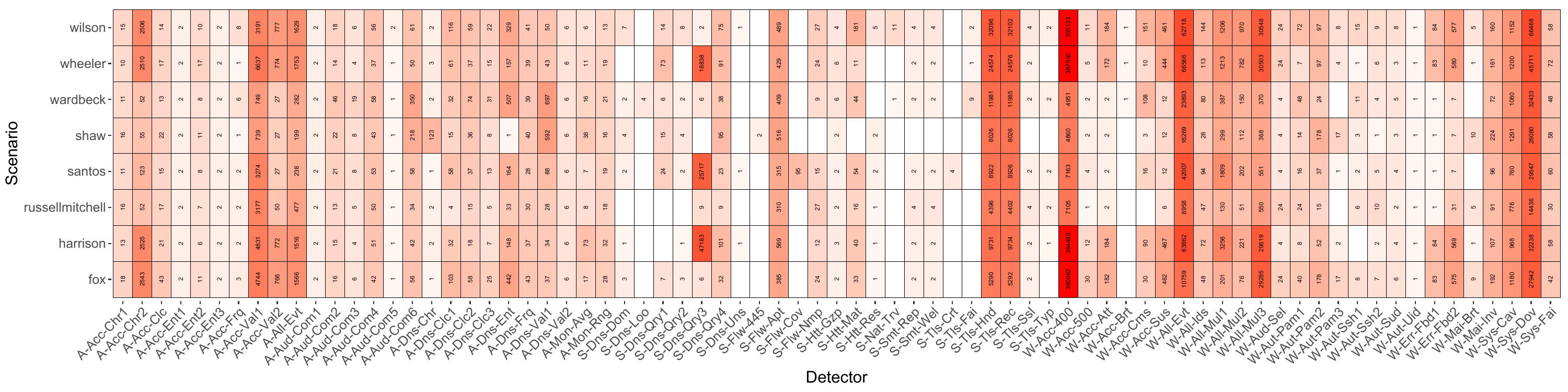}
	\caption{Total number of alert occurrences by detection type in each of the eight scenarios.}
	\label{fig:scenario_heat}
	\vspace{-7pt}
\end{figure*}

The plots in the previous section display overall distributions and patterns of alerts, but make it difficult to compare frequencies of alerts as many symbols overlap. We therefore also count the numbers of alerts reported by each detector for a quantitative comparison. Figure \ref{fig:scenario_heat} shows a heatmap of reported alerts in scenarios, where darker colors indicate higher alert frequencies and the exact numbers are written in the respective cells. This plot allows to differentiate the four scenarios with extensive scanning (\textit{fox}, \textit{harrison}, \textit{wheeler}, \textit{wilson}) from those with basic scanning (\textit{russellmitchell}, \textit{santos}, \textit{shaw}, \textit{wardbeck}) as the latter have significantly less alerts reported by detectors such as \textit{W-Acc-400}, \textit{W-Err-Fbd2}, \textit{A-Acc-Chr2}, etc. The heatmap also shows that there are not only significantly more alerts reported by \textit{S-Dns-Qry3} in the \textit{harrison} and \textit{santos} scenarios as explained in the previous section, but also in the \textit{wheeler} scenario as it uses the ``.biz'' top-level-domain for the network. 

\section{Detector Prioritization} \label{prioritization}

This section proposes a scoring scheme to prioritize detectors for the purpose of filtering alerts with low relevance or false positives.

\subsection{Alert Rates} \label{rates}

As visible in Fig. \ref{fig:datasets}, there is a significant number of false positive alerts in our data set. Such a situation would likely not be tolerable in practice as the large number of alerts puts a high workload on operators. Beside practical issues, alerts reported by detectors that are mainly responsible for false positives are also problematic for the purpose of multi-step attack analysis, because even though they are not related to the attacks, some of them interfere with true positive alerts and thereby hinder identification of relevant alerts and interpreting their context of occurrence. Clearly, the focus of analysts should lie on alerts that are capable of detecting one or more attacks steps but do not appear during normal operation.

To be able to differentiate alerts with low and high relevance, we count the number of alerts during each attack phase and additionally define a time window of normal operation to obtain a baseline of average alerting frequencies for each detector. To this end we leverage the attack times provided in the AIT-LDSv2 and specify a test phase of 5 hours roughly at the same time of day as the multi-step attack but one day earlier. This selection ensures that the test phase does not overlap with the training phase of the AMiner and that the average alert occurrence frequencies are comparable as both test phase and attack phases are affected by a similar background noise of normal activity. 

\begin{figure*}
	\centering
	\includegraphics[width=1\textwidth]{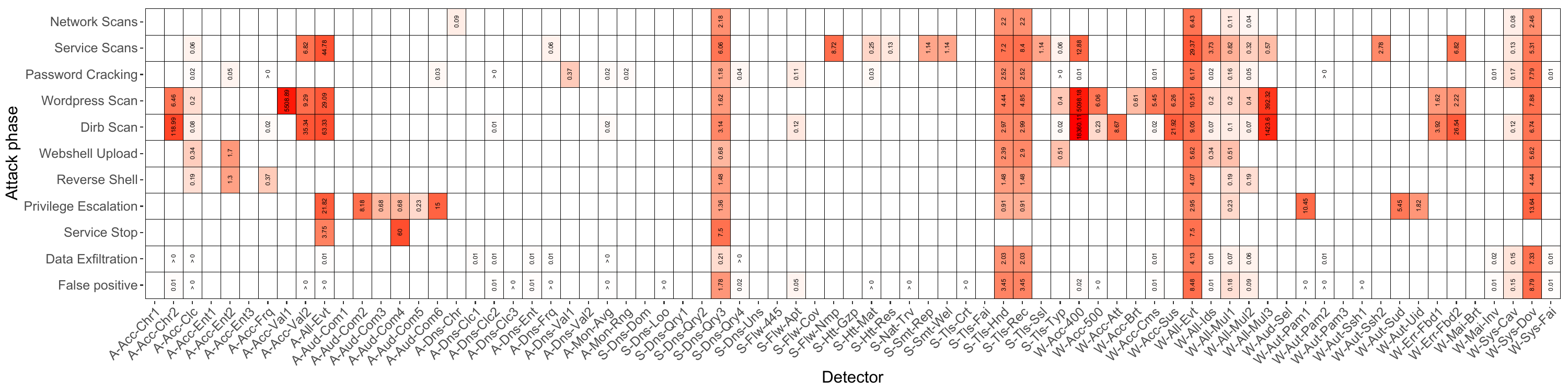}
	\caption{Average alert occurrences per minute by detection type during attack phases and normal operation.}
	\label{fig:attack_heat}
	\vspace{-7pt}
\end{figure*}

Figure \ref{fig:attack_heat} provides a heatmap of the alert rates, i.e., alert occurrences per minute accumulated over all scenarios and for each detector, where very infrequent rates of less than $0.01$ alerts per minute are depicted as $>0$ and empty cells indicate that no alerts are raised by the detector in the respective time interval. We emphasize that counting is only based on the alert timestamp, i.e., the counts for a detector represent how many alerts it produces in the respective time interval without validating that the alert is actually a direct consequence of an attack. The plot already gives a good idea about which detectors are more useful than others. For example, we consider \textit{W-Acc-Att} as a highly valuable detector for producing relevant and specific alerts as it reports on average more than 8 alerts per minute during the Dirb scan and not a single alert during any other observed time interval. Detectors such as \textit{A-Acc-Val2} do involve few false alerts, however, we consider them useful as they also report comparatively high numbers of true positive alerts during specific attack phases. On the contrary, detectors such as \textit{S-Dns-Loo} that only report false positives, \textit{S-Tls-Hnd} that involve similar alert rates for phases of attack and normal behavior, or \textit{W-Mai-Brt} that do not report any alerts in the observed time intervals at all, do not contribute to attack detection in our scenarios. We emphasize that these detectors may be useful for other attack cases and that our analysis only focuses on detection of the attacks involved in the AIT-LDSv2. In the following section, we leverage alert rates to compute scores for detectors that enable ranking. 

\subsection{Detector Scores}

Based on the insights from the previous section, we propose to compute quantitative scores for detectors based on how many alerts they report during attacks and intervals of normal activity. To this end, we compare the number of alerts reported by a detector $D$ during the time interval $\Delta_A$ of attack phase $A$ with the number of false positives in the observed test interval of length $\Delta_T$, weigh them by the duration of the intervals to compensate for the fact that longer time intervals are more likely to contain randomly occurring false positives, and average their ratio for each scenario $S$. Formally, Eq. \ref{rob_score} describes how this score is computed, where the number sign (\#) yields the set size and $\mathcal{A}_{D, S}$ is the set of all alerts from detector $D$ in scenario $S$. We refer to this metric as the robustness score $s_{rob}(A, D)$, because it assesses how robust a detector $D$ reports alerts for attack phase $A$ with respect to false positives caused by normal behavior noise. The robustness score lies in the interval $\left[ 0, 1 \right] $, where values closer to $1$ indicate that the detector reports significantly more alerts during attack phases than normal behavior and values closer to $0$ indicate that false positives are dominating.
\begin{equation} \label{rob_score}
	s_{rob}(A, D) = \frac{1}{\#S} \sum_{S} \left( 1 - min\left( 1, \frac{\#(\mathcal{A}_{D, S} \textrm{ in } \Delta_{T, S})}{\#(\mathcal{A}_{D, S} \textrm{ in } \Delta_{A, S})} \cdot \frac{\Delta_{A, S}}{\Delta_{T, S}} \right) \right)
\end{equation}
Another important factor when it comes to assessing the relevance of detectors is the ability to recognize a specific attack phase independent from attack parameters and other influences from the technical environment or system utilization. In our situation, this means an attack should optimally be detected in a robust way across all scenarios. We therefore multiply the robustness score $s_{rob}(A, D)$ with the ratio of scenarios where detector $D$ triggered at least one alert in the respective interval $\Delta_A$. Equation \ref{det_score} formally describes how to compute the detection score $s_{det}(D)$. Note that we seek the maximum detection score achieved for any attack as we consider an accurate and robust detection of a single attack step as sufficient for a detector to be regarded as relevant; capabilities of a single detector to detect multiple attack phases are beneficial, but not essential. Again, the detection score lies in the interval $\left[ 0, 1 \right] $, with higher values indicating superior detection performance.
\begin{equation} \label{det_score}
	s_{det}(D) = \max_A \left( s_{rob}(A, D) \cdot \frac{\#(S: A \in S \land \#(\mathcal{A}_{D, S} \textrm{ in } \Delta_{A, S}) > 0)}{\#(S : A \in S)} \right)
\end{equation}
Table \ref{tab:overview} shows the number of scenarios where a specific detector reports at least one alert during each of the attack phases. For example, the first row shows that \textit{W-All-Mul3} reports alerts for both WordPress and Dirb Scans in every single scenario; given that the detector does not report any false alerts in the test interval, it can thus be considered highly relevant. The detector also reports alerts for the service scans, however, only manages to do so in five out of eight scenarios and thus falls short to detectors such as \textit{W-Aut-Ssh2} that appears better suited to detect that attack phase.

The two rightmost columns of the table state the computed robustness and detection scores for all detectors. Note that we sorted the table by detection score and omit detectors with a score of $0$ as they do not detect any attacks at all. Generally, high ranked detectors are hardly affected by false positives and may even detect more than one attack phase, while low ranked detectors are less robust, often fail to detect attacks consistently across scenarios, or both. Consider \textit{A-Aud-Com2} as an example, which detects the privilege escalation attack phase across all scenarios without triggering any false alerts in the observed test interval; accordingly, the best possible scores of $s_{rob}(\textrm{Priv. Esc.}, \textit{A-Aud-Com2}) = 1$ and $s_{det}(\textit{A-Aud-Com2}) = 1$ are achieved. \textit{A-Mon-Avg} on the other hand is affected by false positives. In the \textit{fox} scenario, the detector reports one false positive in the test phase of $18,000$ seconds, and one true positive as a consequence of the increased CPU resource consumption that takes place during the password cracking attack phase lasting for $2,120$ seconds. This yields a robustness of $1 - \frac{1}{1} \cdot \frac{2,120}{18,000} = 0.88$ in the \textit{fox} scenario; averaged over all scenarios we obtain a robustness score of $s_{rob}(\textrm{Password cracking}, \textit{A-Mon-Avg}) = 0.94$. As shown in Table \ref{tab:overview}, the detector only raises alerts in 6 out of 7 scenarios where the password cracking attack takes place, resulting in a detection score of $s_{det}(\textit{A-Mon-Avg}) = 0.94 \cdot \frac{6}{7} = 0.8$. In comparison, \textit{A-Mon-Rng} does not report any false positives but fails to detect the password cracking attack in one additional scenario, resulting in a slightly lower detection score of $s_{det}(\textit{A-Mon-Rng}) = 0.71$.

\begin{table}
	\renewcommand{\arraystretch}{0.65}
	\tiny
	\caption{Number of scenarios with at least one alert reported during attack phase and derived scores for each detector}
	\label{tab:overview}
	\begin{tabular}{lccccccccccccc}
\textbf{Detector} & \rotatebox[origin=l]{90}{\textbf{Network Scans}} & \rotatebox[origin=l]{90}{\textbf{Service Scans}} & \rotatebox[origin=l]{90}{\textbf{WordPress Scan}} & \rotatebox[origin=l]{90}{\textbf{Dirb Scan}} & \rotatebox[origin=l]{90}{\textbf{Webshell Upload}} & \rotatebox[origin=l]{90}{\textbf{Password Cracking}} & \rotatebox[origin=l]{90}{\textbf{Reverse Shell}} & \rotatebox[origin=l]{90}{\textbf{Privilege Escalation}} & \rotatebox[origin=l]{90}{\textbf{Service Stop}} & \rotatebox[origin=l]{90}{\textbf{Data Exfiltration}} & \rotatebox[origin=l]{90}{\textbf{False positives}} & \rotatebox[origin=l]{90}{\textbf{Robustness Score}} & \rotatebox[origin=l]{90}{\textbf{Detection Score}} \\ \hline
W-All-Mul3 &   & 5 & 8 & 8 &   &   &   &   &   &   &   & 1.0 & 1.0 \\ \hline
W-Acc-Sus &   &   & 6 & 8 &   &   &   &   &   &   &   & 1.0 & 1.0 \\ \hline
W-Acc-Att &   &   &   & 8 &   &   &   &   &   &   &   & 1.0 & 1.0 \\ \hline
W-Err-Fbd2 &   & 5 & 3 & 8 &   &   &   &   &   &   &   & 1.0 & 1.0 \\ \hline
W-Aut-Ssh2 &   & 8 &   &   &   &   &   &   &   &   &   & 1.0 & 1.0 \\ \hline
W-Aut-Uid &   &   &   &   &   &   &   & 8 &   &   &   & 1.0 & 1.0 \\ \hline
W-Aut-Sud &   &   &   &   &   &   &   & 8 &   &   &   & 1.0 & 1.0 \\ \hline
W-Err-Fbd1 &   &   & 8 & 4 &   &   &   &   &   &   &   & 1.0 & 1.0 \\ \hline
A-Aud-Com4 &   &   &   &   &   &   &   & 3 & 8 &   &   & 1.0 & 1.0 \\ \hline
A-Aud-Com2 &   &   &   &   &   &   &   & 8 &   &   &   & 1.0 & 1.0 \\ \hline
A-Aud-Com6 &   &   &   &   &   & 1 &   & 8 &   &   &   & 1.0 & 1.0 \\ \hline
A-Acc-Val1 &   &   & 8 &   &   &   &   &   &   &   &   & 1.0 & 1.0 \\ \hline
A-Acc-Ent2 &   &   &   &   & 8 & 7 & 7 &   &   &   &   & 1.0 & 1.0 \\ \hline
W-Acc-400 &   & 7 & 8 & 8 &   & 1 &   &   &   &   & 4 & 1.0 & 1.0 \\ \hline
A-All-Evt &   & 8 & 8 & 8 &   &   &   & 8 & 1 & 1 & 2 & 1.0 & 1.0 \\ \hline
W-Acc-500 &   &   & 8 & 4 &   &   &   &   &   &   & 1 & 1.0 & 1.0 \\ \hline
A-Acc-Val2 &   & 5 & 8 & 8 &   &   &   &   &   &   & 2 & 1.0 & 1.0 \\ \hline
W-Aut-Pam1 &   &   &   &   &   &   &   & 8 &   &   & 1 & 1.0 & 1.0 \\ \hline
A-Acc-Chr2 &   &   & 8 & 8 &   &   &   &   &   & 1 & 1 & 1.0 & 1.0 \\ \hline
S-Smt-Wel &   & 7 &   &   &   &   &   &   &   &   &   & 1.0 & 0.88 \\ \hline
S-Smt-Rep &   & 7 &   &   &   &   &   &   &   &   &   & 1.0 & 0.88 \\ \hline
S-Flw-Nmp &   & 7 &   &   &   &   &   &   &   &   &   & 1.0 & 0.88 \\ \hline
S-Tls-Ssl &   & 7 &   &   &   &   &   &   &   &   &   & 1.0 & 0.88 \\ \hline
W-All-Ids &   & 7 & 1 & 2 & 2 & 5 &   &   &   & 4 & 6 & 1.0 & 0.87 \\ \hline
A-Mon-Avg &   &   &   & 2 &   & 6 &   &   &   & 1 & 4 & 0.94 & 0.8 \\ \hline
A-Mon-Rng &   &   &   &   &   & 5 &   &   &   &   &   & 1.0 & 0.71 \\ \hline
W-All-Evt & 5 & 7 & 5 & 4 & 5 & 7 & 3 & 2 & 1 & 7 & 8 & 0.8 & 0.7 \\ \hline
W-All-Mul1 & 5 & 6 & 1 & 3 & 3 & 6 & 1 & 1 &   & 5 & 8 & 0.81 & 0.61 \\ \hline
S-Tls-Rec & 5 & 7 & 5 & 4 & 6 & 6 & 3 & 1 &   & 7 & 8 & 0.57 & 0.5 \\ \hline
A-Acc-Clc &   & 1 & 1 & 4 & 2 & 3 & 1 &   &   & 1 & 1 & 0.99 & 0.49 \\ \hline
W-All-Mul2 & 4 & 4 & 2 & 3 &   & 5 & 1 &   &   & 4 & 7 & 0.9 & 0.45 \\ \hline
S-Htt-Mat &   & 1 &   &   &   & 3 &   &   &   &   & 2 & 0.94 & 0.4 \\ \hline
S-Tls-Typ &   & 1 & 2 & 1 & 3 & 1 &   &   &   &   &   & 1.0 & 0.38 \\ \hline
A-Aud-Com3 &   &   &   &   &   &   &   & 3 &   &   &   & 1.0 & 0.38 \\ \hline
W-Acc-Brt &   &   & 3 &   &   &   &   &   &   &   &   & 1.0 & 0.38 \\ \hline
W-Acc-Cms &   &   & 3 & 1 &   & 2 &   &   &   & 4 & 5 & 1.0 & 0.37 \\ \hline
S-Flw-Apt &   &   &   & 1 &   & 3 &   &   &   &   & 8 & 0.82 & 0.35 \\ \hline
W-Mai-Inv &   &   &   &   &   & 1 &   &   &   & 3 & 5 & 0.8 & 0.3 \\ \hline
W-Sys-Fai &   &   &   &   &   & 1 &   &   &   & 3 & 5 & 0.8 & 0.3 \\ \hline
W-Aut-Pam2 &   &   &   &   &   & 1 &   &   &   & 3 & 5 & 0.8 & 0.3 \\ \hline
W-Sys-Dov & 7 & 3 & 5 & 4 & 3 & 6 & 5 & 5 &   & 7 & 8 & 0.46 & 0.29 \\ \hline
S-Tls-Hnd & 5 & 3 & 3 & 4 & 3 & 6 & 3 & 1 &   & 7 & 8 & 0.42 & 0.26 \\ \hline
S-Htt-Res &   & 2 &   &   &   &   &   &   &   &   &   & 1.0 & 0.25 \\ \hline
A-Dns-Clc1 &   &   &   &   &   &   &   &   &   & 2 &   & 1.0 & 0.25 \\ \hline
A-Dns-Frq &   & 1 &   &   &   &   &   &   &   & 2 & 1 & 1.0 & 0.25 \\ \hline
A-Acc-Frq &   &   &   & 2 &   & 1 & 2 &   &   &   &   & 1.0 & 0.25 \\ \hline
W-Sys-Cav & 1 & 1 &   & 2 &   & 7 &   &   &   & 8 & 8 & 0.24 & 0.24 \\ \hline
S-Dns-Qry4 &   &   &   &   &   & 2 &   &   &   & 2 & 6 & 0.85 & 0.24 \\ \hline
A-Dns-Clc2 &   &   &   & 1 &   & 1 &   &   &   & 3 & 5 & 0.5 & 0.19 \\ \hline
A-Dns-Val1 &   &   &   &   &   & 1 &   &   &   &   &   & 1.0 & 0.14 \\ \hline
A-Dns-Chr & 1 &   &   &   &   &   &   &   &   &   &   & 1.0 & 0.12 \\ \hline
A-Aud-Com5 &   &   &   &   &   &   &   & 1 &   &   &   & 1.0 & 0.12 \\ \hline
S-Dns-Qry3 & 2 & 1 & 1 & 2 & 1 & 1 & 2 & 1 & 1 & 1 & 2 & 0.88 & 0.11 \\ \hline
A-Dns-Ent &   &   &   &   &   &   &   &   &   & 1 & 2 & 0.63 & 0.08 \\ \hline
	\end{tabular}
\vspace{-7pt}
\end{table}

\subsection{Attack Timelines}

Utilizing the scores computed in the previous section allows to identify relevant detectors and select only those alerts originating from them for analysis of multi-step attacks. In particular, we select $s_{det} > 0.7$ as a cutoff threshold for our further analyses as all detectors with lower scores appear to be unreliable or designed for broad rather than attack-specific detection according to Table \ref{tab:overview}.

\begin{figure*}
	\centering
	\includegraphics[width=1\textwidth]{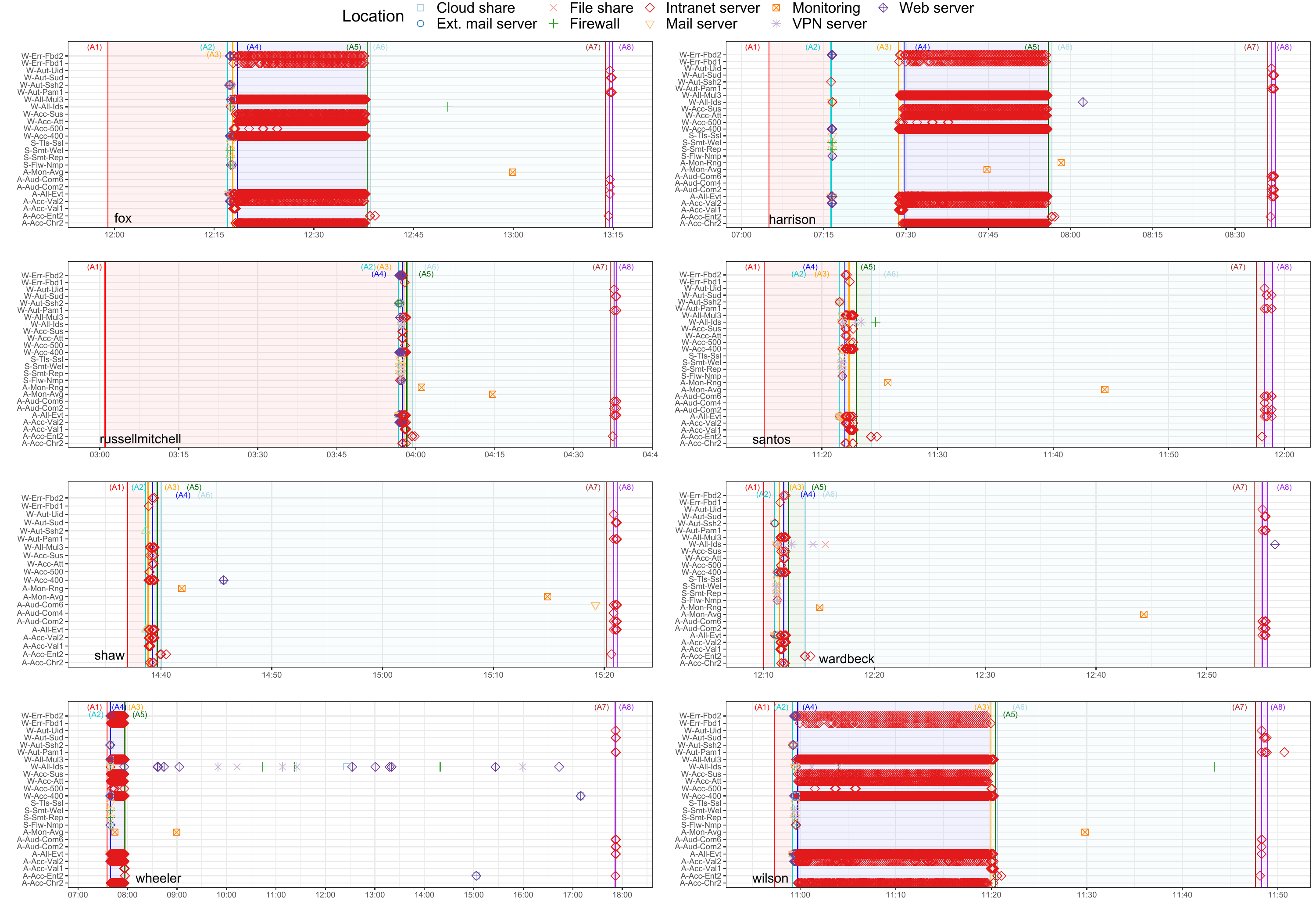}
	\caption{Timelines of alerts reported by highly ranked detectors during multi-step attack. Shaded intervals indicate network scans (A1/red), service scans (A2/cyan), WordPress scan (A3/yellow) Dirb scan (A4/blue), webshell upload and command execution (A5/green), password cracking (A6/light blue), reverse shell (A7/brown), and privilege escalation (A8/purple).}
	\label{fig:alerts}
	\vspace{-7pt}
\end{figure*}

Figure \ref{fig:alerts} visualizes the alerts of the resulting selection of 26 detectors, where the horizontal axis is limited on the time interval of the multi-step attack in the respective scenario. Similar patterns of alert appearances are visible across all scenarios and hardly any false positives remain, except for \textit{W-All-Ids} that produces several alerts in the time interval between the webshell and reverse shell upload of the wheeler scenario (note that this scenario lacks the password cracking attack). This view on the multi-step attack also shows that the same attack phases vary across the scenarios; this concerns the lengths of phases and also attack parameters such as extensiveness of scans, which results in higher numbers of alerts in \textit{fox}, \textit{harrison}, \textit{wheeler}, and \textit{wilson} in comparison to the other scenarios. In addition to variations in single attack phases, the order of Dirb (A4) and WPScan (A5) also changes across scenarios. These subsets of our alert data set are the input to the multi-step analysis methods explored in the following section.

\section{Multi-Step Attack Analysis} \label{analysis}

This section covers an illustrative application of our alert data set using two frameworks for multi-step attack analysis (cf. Sect. \ref{related}). We specifically select these two approaches, because they are among the few scientific frameworks available as open-source, have only recently been published, and are capable of showing two different goals in the research are of multi-step attack analysis - aggregation of alerts into meta-alerts and derivation of attack graphs.

\subsection{Alert Aggregation} \label{aggregation}

The main goal of alert aggregation is to identify repeating patterns of individual attack steps or fine-grained actions that make up multi-step attacks, and generate meta-alerts for each of these patterns. Thereby, meta-alerts are abstract representations of specific activities that are generated by merging two or more alerts related by some logical connection, e.g., similarity or co-occurrence \cite{valeur2004comprehensive}.

To this end we select the AECID-Alert-Aggregation framework \cite{landauer2022dealing} that is publicly available as open-source software on GitHub\footnote{AECID-Alert-Aggregation repository, \url{https://github.com/ait-aecid/aecid-alert-aggregation} (accessed 2023-08-10)}. The approach is designed as an incremental procedure that first groups alerts that occur close in time based on the assumption that co-occurring alerts are possibly related to each other, then measures the similarity of these alert groups by comparing alert attributes, alert frequencies, and sequential patterns in alert occurrences, and eventually merges groups that achieve high similarities. The meta-alerts resulting from this continuous merging strategy are in the same format as the alerts themselves (i.e., JSON format) but the fields and corresponding values are adapted, extended, or removed to correspond to the majority of groups associated with the meta-alert. For example, an attribute of an alert that appears in all groups with different values may be replaced with a wildcard in the meta-alert to indicate that its exact value is irrelevant for identifying the attack, while other attributes with coinciding values across groups may be added to the meta-alert as is. 

For our experiments we feed the filtered alerts from the 26 top ranked detectors according to our prioritization technique (cf. Sect. \ref{prioritization}) that occur in any of the known attack phases into the aggregation framework. We manually investigate alert occurrences in the multi-step attack to find a suitable value for the parameter that is referred to as interval time by the original authors. Since this parameter is crucial for grouping related alerts as it specifies the minimum time without any alert occurrences between any two groups, we select an interval time of 2 seconds as this appears large enough to group alerts within the same attack phase but short enough to avoid grouping alerts of distinct phases. Across all scenarios we obtain 150 groups of alerts that are formed within the attack phases.

\begin{figure*}
	\centering
	\includegraphics[width=1\textwidth]{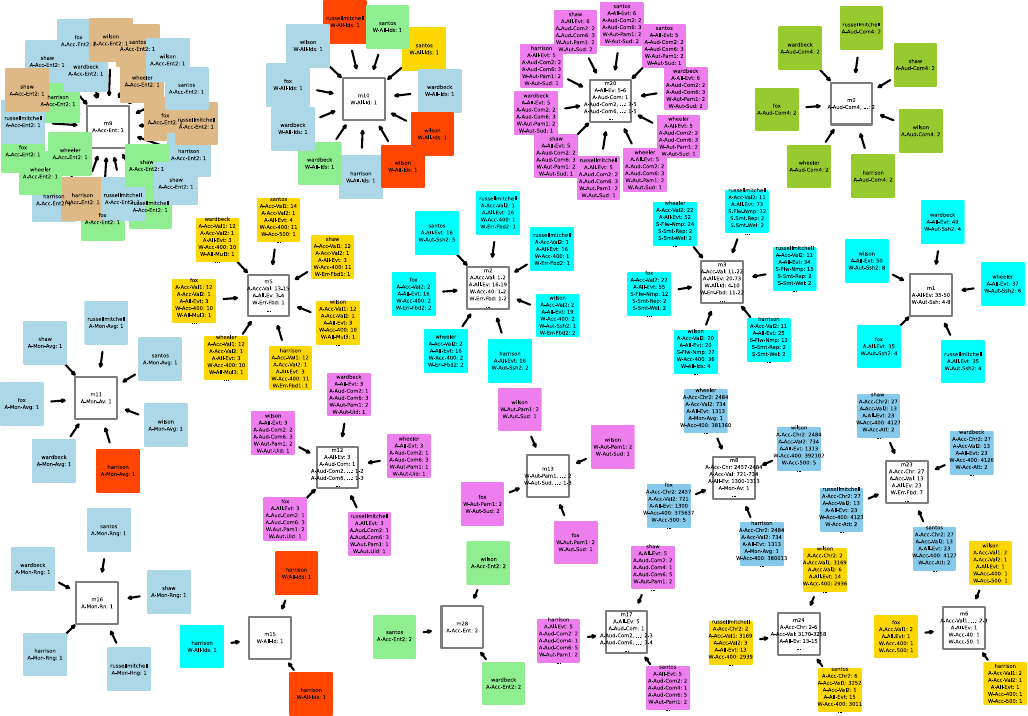}
	\caption{Meta-alerts and alert groups generated by the AECID-alert-aggregation framework for service scans (cyan), WordPress scan (yellow), Dirb scan (blue), webshell upload (green), password cracking (light blue), reverse shell (brown), privilege escalation (purple), service stop (dark green), and data exfiltration (red) attack phase.}
	\label{fig:aecid}
\end{figure*}

There are two parameters that specify the minimum similarity thresholds for merging groups and alerts, which we set to 0.55 and 0.5 after empirically validating the results. With these settings the aggregation framework merges the alert groups into 42 meta-alerts. Several of these meta-alerts only correspond to a single group of alerts; the main focus of our experiment, however, lies on those meta-alerts that represent the same or similar attack phases across multiple scenarios. Figure \ref{fig:aecid} therefore visualizes all meta-alerts (white squares) with at least three corresponding groups that we color-code by attack phase. In each group we print the scenario and a (possibly truncated) list of involved alerts, including their frequencies. The meta-alerts comprise an unique identifier and also a (possibly truncated) list of merged alerts. Overall, the visualization indicates that most alert groups belonging to the same stages of the multi-step attack are correctly merged together. For example, meta-alert \textit{m0} corresponding to the service stop phase (green squares) is merged from groups containing similar alerts from seven out of eight scenarios. Since the Dirb scan is executed in extensive and basic mode in different scenarios (cf. Sect. \ref{counts}), two distinct meta-alerts \textit{m8} and \textit{m23} form that correspond to each of the execution modes. Moreover, the plot reveals that some attack phases comprise multiple sub-steps that actually need more fine-granular labels, e.g., the service scans (cyan squares) yield three distinct meta-alerts. These findings align with the situation faced by the original authors of the approach \cite{landauer2022dealing}. While most meta-alerts only contain groups that belong to the same attack phase, the figure shows that groups containing only a single alert appear more difficult to cluster correctly; specifically, this concerns meta-alerts \textit{m9} and \textit{m10}. The reason for this is that the same detector raises alerts for multiple attack phases and a single alert is thus not specific enough to act as a unique identifier for some attack phase. Overall, these results suggest that our alert data set is suitable to develop and evaluate alert aggregation approaches, because alert patterns of some attack phases are more difficult to cluster and merge than others and the generation of a set of meta-alerts that subsumes all alert groups remains a challenge.

\subsection{Attack Graph Mining} \label{graph}

Alert aggregation as illustrated in the previous section is usually a static approach to multi-step attack analysis that may consider alert sequences for the purpose of similarity computation but puts less focus on the recreation of the sequential execution stages of attacks, such as linking meta-alerts into chains. On the contrary, attack graphs specifically aim to visually summarize attack strategies by connecting related steps that attackers need to take to achieve their goals. Generation of such graphs is often a tedious process that requires expert knowledge; however, there are also attempts to ease this task through automation. One of them is SAGE \cite{nadeem2021sage}, an open-source tool that is available on GitHub\footnote{SAGE repository, \url{https://github.com/tudelft-cda-lab/SAGE} (accessed 2023-08-10)} and enables alert-driven attack graph extraction from raw intrusion alert sequences.

\begin{figure*}
	\centering
	\includegraphics[width=\textwidth]{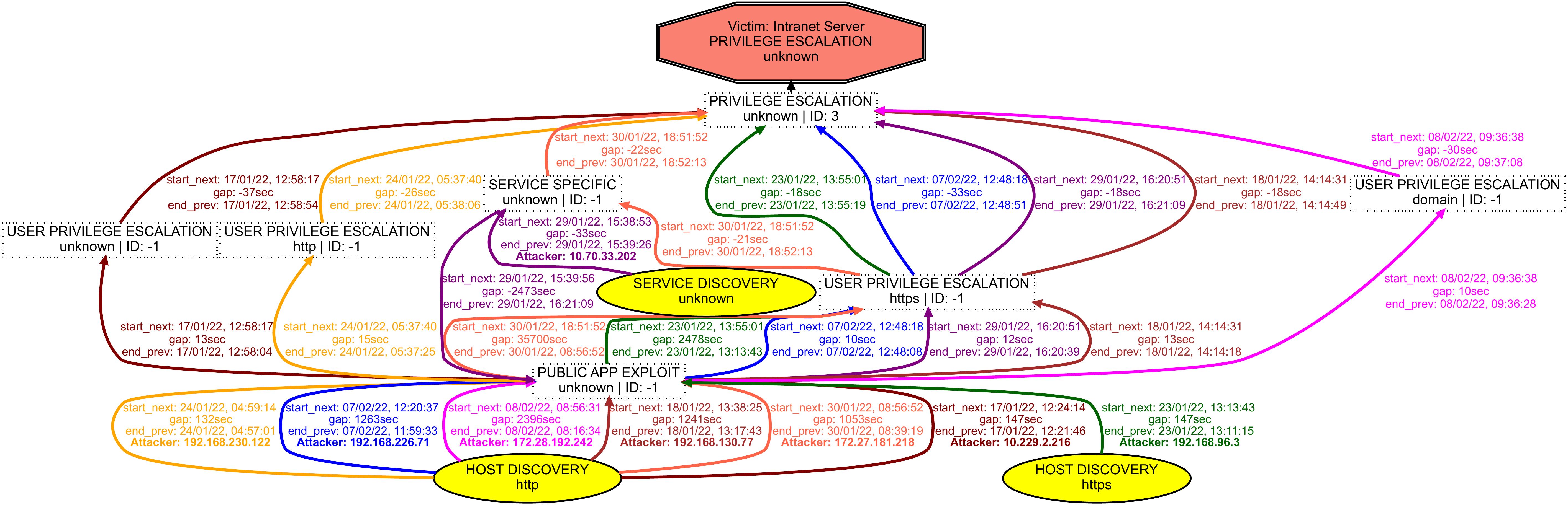}
	\caption{Attack graph generated by the SAGE framework showing the steps in which attackers conducted the multi-step attack in fox (maroon), harrison (pink), russellmitchell (yellow), santos (brown), shaw (purple), wardbeck (green), wheeler (orange), and wilson (blue) scenario.}
	\label{fig:sage}
\end{figure*}

The framework relies on a manually crafted mapping of intrusion detection alerts to some higher-level attack stages, such as scanning, exploit, or privilege escalation. In particular, the authors rely on the categories provided by the Action-Intent Framework \cite{moskal2020framework}, which draw away from technical details of alert signatures and focus on goals and strategies of attackers when classifying alerts. As an initial step, SAGE filters irrelevant alerts, in particular, alerts that relate to non malicious activities (according to the mapping) and repetitions of alerts within short time intervals. The remaining alerts and their corresponding attack stages are then arranged into so-called episodes of attacker behavior, i.e., short sequences of alerts that relate to distinct actions similar to alert groups described in Sect. \ref{aggregation}. SAGE then leverages FlexFringe \cite{verwer2022flexfringe}, an open-source framework to generate probabilistic automata for software behavior from logs, to merge sequential episodes and subsequently create graphs. While the resulting graphs comprise a single end node representing the goal of the attackers, they may comprise multiple start nodes to represent all the possible paths attackers take to achieve their goals. Finally, SAGE also leverages alert attributes to enrich the resulting graph, for example, port numbers are extracted to identify the services targeted by a specific attack step. 

The original authors of SAGE assume that several attacker teams operate on the same infrastructure and thus pursue the generation of separate attack graphs for every victim system to understand how different teams attack the same component. Since the infrastructures in our scenarios are almost identical in terms of available systems but isolated from each other, we therefore modify the data so that the attackers appear to target the same systems. To this end we overwrite all IP addresses of victim components to coincide with those of their counterparts across all scenarios. Moreover, SAGE relies on two crucial parameters that set the time windows to filter redundant alerts and aggregate alerts into episodes, which we set to 2 seconds corresponding to the time interval we used for grouping during alert aggregation (cf. Sect. \ref{aggregation}), and 2 hours to ensure that attack chains are not interrupted, respectively. Figure \ref{fig:sage} displays the resulting attack graph for the intranet server, which is the primary target of the multi-step attack. Attackers are color-coded according to their corresponding scenarios and arrows indicate their progression in compromising the intranet server. The graph shows that the steps taken by attackers are similar across all scenarios, starting with scanning activities (``host discovery'' and ``service discovery''), continuing with the exploit of the WordPress platform (``public app exploit''), until access is gained via the webshell (``user privilege escalation'') and eventually privileges are escalated (``privilege escalation''). Note that some attack phases are duplicated as different ports are involved. Moreover, attacks such as the password cracking phase are missing from the graph as the corresponding alerts do not occur sufficiently consistent across scenarios. Overall, the attack graph provides a compact and high-level overview of common patterns and dependencies of attack phases but does not allow to differentiate attacks on a fine-granular level, e.g., WordPress and Dirb scans are merged into a single state. Due to its challenging mix of similar alert patterns and diverse attack manifestations, our alert data set offers a suitable basis for the development and evaluation of new algorithms for attack graph mining.

\section{Discussion} \label{discussion}

We generated the alert data set that is introduced in this paper with several use-cases and requirements in mind; specifically, we aimed to resolve issues with existing alert data sets and designed the data set to address challenges in the research domain of multi-step attack analysis \cite{navarro2018systematic, kotenko2022systematic, husak2018survey}. The main characteristics of the alert data set that differentiate it - to the best of our knowledge - from data sets commonly used by researchers are as follows. (i) The data set is publicly available and thus facilitates reproducibility of evaluations. We publish configuration files of deployed IDSs so that others are able to replicate the data set or produce variants of it by changing configuration parameters, e.g., adapting detectors (see link in Sect. \ref{intro}). (ii) The data set involves several multi-step attacks carried out independently and with variations in eight different scenarios (cf. Sect. \ref{aitlds}), enabling aggregation and meta-alert generation (cf. Sect. \ref{analysis}). (iii) Alerts are generated from three different IDSs with heterogeneous detection techniques that analyze multiple sources, involve diverse alert formats (cf. Sect. \ref{ids}), and stem from all relevant components in the network (cf. Sect. \ref{generation}). (iv) Since the alerts are generated from a synthetic and labeled log data set, we are also able to provide labels for attack phases based on alert occurrences (cf. Sects. \ref{aitlds} and \ref{timelines}). (v) The data set also involves a high number of false alerts and is thus suitable for evaluation of alert prioritization and filtering (cf. Sects. \ref{timelines} and \ref{prioritization}).

\begin{table*}
	\setlength{\tabcolsep}{4pt}
	\scriptsize
	\caption{Effect of filtering on the numbers of alerts in scenarios and average reduction rates}
	\label{tab:results}
	\begin{tabular}{lccccccccc}
		\textbf{Alerts} & \textbf{fox} & \textbf{harrison} & \textbf{russellmitchell} & \textbf{santos} & \textbf{shaw} & \textbf{wardbeck} & \textbf{wheeler} & \textbf{wilson} & \textbf{Avg. reduction rate} \\ \hline
		All & 473,104  & 593,948 & 45,544 & 130,779 & 70,782  &  91,257 & 616,161  & 634,246  & - \\ \hline
		Filtered by prioritization & 420,600 (11.10\%)  & 425,392 (28.38\%)  & 11,705 (74.30\%) & 11,709 (91.05\%) & 6,667 (90.58\%)  & 7,107 (92.21\%) & 431,319 (30.00\%)  & 435,538 (31.33\%)  & 56.12\% \\ \hline
		In attack phases & 421,653 (10.88\%)  & 431,492 (27.35\%) & 12,015 (73.62\%) & 13,004 (90.06\%) & 6,935 (90.20\%)  & 7,040 (92.29\%)  &  432,334 (29.83\%) & 440,108 (30.61\%) & 55.6\% \\ \hline
		Filtered and in attack phases &  420,112 (11.20\%) & 424,974 (28.45\%)  & 11,230 (75.34\%) & 11,217 (91.42\%) & 6,065 (91.43\%)  & 6,213 (93.19\%) & 430,737 (30.09\%) & 434,952 (31.42\%)  & 56.57\% \\ \hline \hline
		SAGE & 5,755 (98.63\%) & 6,515 (98.47\%) & 383 (96.59\%) & 238 (97.88\%) & 175 (97.11\%) & 210 (96.62\%) & 6,785 (98.42\%) & 8,209 (98.11\%) & 97.73\% \\ \hline
		AECID-Alert-Aggregation & 167 (99.96\%) & 167 (99.96\%) & 167 (98.51\%) & 167 (98.51\%) & 167 (97.25\%) & 167 (97.31\%) & 167 (99.96\%) & 167 (99.96\%) & 98.93\% \\ \hline
	\end{tabular}
\vspace{-7pt}
\end{table*}

We see several interesting research opportunities enabled by our data set. For example, we showed in this paper how prioritization may be applied on the data to filter alerts and identify relevant detectors. Thereby, our prioritization technique that assigns scores to each detector relies on the labels of alerts to differentiate between time intervals corresponding to benign and attacker behavior, which only works in a supervised setting. In real-world applications, however, labels for system activities are generally not available or reliable, which requires unsupervised approaches or possibly semi-supervised training phases where only normal activity occurs on the systems. Another problem that could hinder practical application of our prioritization technique is that a higher number of false positives is tolerated for detectors that report many true positives, which may not be accepted in practice. While the scores we compute for detectors could be used as a basis for comparison, these problems demand the exploration of new methods for alert prioritization, which is a task that we leave for future work.

One of the key metrics used by researchers to compare approaches on alert filtering and aggregation is the reduction rate, which measures what percentage of alerts does not need to be reviewed by human operators. Table \ref{tab:results} compares reduction rates that we achieved on our alert data set using the frameworks explored in this paper. The first row shows the total number of alerts generated in each of the eight scenarios. The second row shows how many alerts remain after applying our prioritization technique, i.e., only considering alerts from detectors with a detection score of more than $0.7$, as well as the reduction rate with respect to the total number of alerts in brackets. The third row then shows the numbers of alerts from any detectors that occur within one of the attack phases. Applying both filtering techniques in combination, i.e., only considering alerts by relevant detectors that occur within attack phases, yields almost the same numbers as before in each scenario and an average reduction rate of 56.57\% across all scenarios. We consider this as a validation that our prioritization was able to select those detectors as relevant that produce many alerts related to attacks while filtering false positives. Since these are the alerts we feed into the aggregation approaches from Sect. \ref{analysis}, we compute their reduction rates based on these counts. SAGE removes duplicates of alerts within time windows and achieves an average reduction rate of 97.73\%. AECID-Alert-Aggregation on the other hand finds similar groups of alerts across all scenarios and merges them to meta-alerts. Enabled by the fact that alerts have frequency information attached to them that specifies the expected range of occurrences (cf. Sect. \ref{aggregation}), the resulting meta-alerts comprise a total of only 167 distinct alerts, corresponding to an average reduction rate of 98.93\%.

We illustrate that both SAGE and AECID-Alert-Aggregation yield interesting and useful results when applied on our data set; however, we also want to point out some ideas for future work that could further improve these concepts. While the graph extraction of SAGE itself is unsupervised, it hinges on a mapping of detector signatures to high-level attack phases that needs to be created through expert knowledge. This may be difficult in practice as an exhaustive list of signatures is not necessarily known and may change over time. Even more problematic is the fact that some alerts possibly fit into more than one stage of the kill chain. In particular, alerts from anomaly-based IDSs are often too generic to be mapped to a specific attack, for example, log events occurring with unusual frequencies could be related to basic scans, brute-force attacks, or some activity related to data exfiltration as shown in Sect. \ref{timelines}. The AECID-Alert-Aggregation on the other hand is fully unsupervised, but does not show the progression of attack steps. We thus propose to combine the advantages of both methods and apply SAGE's algorithm for attack graph generation on the sequences of meta-alerts identified by AECID-Alert-Aggregation. This could allow to derive attack graphs with high technical detail regarding the involved alerts in different stages, which could in turn enable automatic recognition of attacks that follow similar attack stages, attribution of observed multi-step attacks to certain adversarial actors, as well as prediction of subsequent attack phases.

We also share some insights regarding challenges that need to be considered for the generation of new alert data sets. Most of all, alerts are obviously heavily dependent on the selection and configuration of IDSs. Unfortunately, designing a suitable setup of IDSs is non-trivial since configurations are likely very diverse in real-world scenarios and specifically the configuration of anomaly-based IDSs highly depends on expert knowledge about the monitored systems. This also concerns the time used to train detectors utilizing machine learning techniques. Additional detectors with advanced analysis techniques as well as a more extensive set of detection rules could generate more distinct alerts and thus further improve the results of multi-step attack analysis. Overall, we believe that IDS selection and configuration would benefit from a structured analysis of attack manifestations in log data that investigates how and in what sources different attack techniques leave traces suitable for detection. In addition, we also argue that alert data sets comprising overlapping attack phases caused by one or multiple adversaries launching several attacks at the same time could result in more challenging alert data sets with relevance for advanced analysis of multi-step attacks. We leave these tasks for future work.

\section{Conclusion} \label{conclusion}

In this paper we describe a novel alert data set that we specifically generate for the purpose of evaluating approaches in the research domain of multi-step attack analysis, such as detection and prediction of attack stages. We collect the data set by forensically analyzing the AIT-LDSv2, a publicly available collection of network traffic and log data sets, and collecting alerts with three different intrusion detection systems, namely Suricata, Wazuh, and AMiner, to generate more than 2.6 million alerts with 93 distinct detector signatures. The data set is designed to overcome prevalent issues with existing data sets by providing alerts from modern and heterogeneous detectors monitoring diverse data sources for traces of relevant and fitting attack steps. As we show in this paper, the properties of the data set make it a promising basis for future research endeavors. Specifically, the presence of alerts with diverse relevance for detection as well as false positive alerts facilitate filtering and prioritization techniques. Since the alerts origin from eight separate environments where the attack steps are executed with variations, the data set also enables generation of meta-alerts and attack graphs. We foresee to use the data set to develop and evaluate approaches for attack pattern recognition that combine the advantages of meta-alerts and attack graphs.

\begin{acks}
This work was partly funded by the European Defence Fund (EDF) project AInception (101103385) and the FFG project PRESENT (FO999899544).
\end{acks}

\bibliographystyle{ACM-Reference-Format}
\bibliography{acmart}


\end{document}